%% file: main.tex
\documentclass[acmsmall]{acmart}
\usepackage{graphicx} 
\usepackage{soul}
\usepackage{enumitem}

\begin{document}

\setcopyright{acmlicensed}
\acmJournal{PACMHCI}
\acmYear{2025} \acmVolume{9} \acmNumber{2} \acmArticle{CSCW117} \acmMonth{4}\acmDOI{10.1145/3711015}

\title[Interrogating AI]{Interrogating AI: Characterizing Emergent Playful Interactions with ChatGPT}

\author{Mohammad Ronagh Nikghalb}
\email{mohammad-ronagh.nikghalb@polymtl.ca}
\affiliation{%
  \institution{Polytechnique Montreal}
  \city{Montreal}
  \state{QC}
  \country{Canada}
}
\author{Jinghui Cheng}
\email{jinghui.cheng@polymtl.ca}
\affiliation{%
  \institution{Polytechnique Montreal}
  \city{Montreal}
  \state{QC}
  \country{Canada}
}

\begin{abstract}
In an era of AI's growing capabilities and influences, recent advancements are reshaping HCI and CSCW's view of AI. Playful interactions emerged as an important way for users to make sense of the ever-changing AI technologies, yet remained underexamined. We target this gap by investigating playful interactions exhibited by users of a popular AI technology, ChatGPT. Through a thematic analysis of 372 user-generated posts on the ChatGPT subreddit, we found that more than half (54\%) of user discourse revolved around playful interactions. The analysis further allowed us to construct a preliminary framework to describe these interactions, categorizing them into six types: reflecting, jesting, imitating, challenging, tricking, and contriving; each included sub-categories. This study contributes to HCI and CSCW by identifying the diverse ways users engage in playful interactions with AI. It examines how these interactions can help users understand AI's agency, shape human-AI relationships, and provide insights for designing AI systems.
\end{abstract}



\maketitle

\input{s_introduction}
\input{s_relatedwork}
\input{s_methods}
\input{s_results}
\input{s_discussion}
\input{s_conclusion}

\begin{acks}
We would like to thank Arman Jafarnezhad for his invaluable support for data extraction. We also thank the anonymous reviewers for helping us improve the paper. This work is partially supported by the Canada Research Chairs program (CRC-2021-00076) and the Natural Sciences and Engineering Research Council of Canada (RGPIN-2018-04470).
\end{acks}

\bibliographystyle{ACM-Reference-Format}
\bibliography{reference}

\received{January 2024}
\received[revised]{July 2024}
\received[accepted]{October 2024}

\end{document}

%% file: s_introduction.tex
\section{Introduction}
Undeniably, artificial intelligence (AI) is becoming increasingly powerful and exerts a greater impact on individuals' lives and our society as a whole. Throughout history, we as humans have continuously navigated our interactions and relationships with such technological advancements. These explorations and reflections are evident over time in various fictional narratives, such as novels, films, and video games (e.g., \textit{I, Robot}; \textit{Neuromancer}; \textit{Blade Runner}; \textit{Her}; \textit{Detroit: Become Human}; among many others). However, it was not until recently that the impact of AI became disturbingly tangible and alarmingly real, which triggered discussion at different levels of society (e.g.,~\cite{AIRisk2023}).

This impact is manifested not only in the rapid proliferation of AI technologies and their expanding capabilities but also in the type of AI currently being created and researched. One influential paradigm of understanding the human-AI interaction in HCI and CSCW is centered around the notion of powerful tools~\cite{Farooq2017, Vaithilingam2024}, largely influenced by the Activity Theory~\cite{Kaptelinin2012}. However, the most recent advances in AI technology, such as technologies leveraging generative models and large language models (LLMs), appeared to have transcended the notion of powerful tools, particularly since they innately have an interactive component but are not created with a specific practical goal in mind. While it remains a crucial task for researchers to explore ways to effectively design with such technologies to deliver practical benefits to users (e.g., through investigating both how designers use AI~\cite{Yildirim2022} and how to leverage AI for design~\cite{Mozaffari2022, Shi2023}), it is imperative to recognize that the implications of these technologies extend beyond practical user interactions.

Nowadays, people do not only ``\textit{use}'' AI technologies, but also ``\textit{play}'' with them~\cite{Liapis2023, playerAIint}. Those interactions are often not predefined by designers or developers. Instead, those interactions, along with the associated experiences, emerge through the process of negotiation and exploration itself. This behavior is similar to what HCI researchers have termed as \textit{appropriation}~\cite{designApprop} and \textit{re-purposing}~\cite{interactionKnowledge, techreason}. While these works recommend designing systems that support user adaptations, as well as understanding and leveraging users' prior knowledge and technical reasoning to improve the design of digital tools, we argue that interactions with recent AI tools are different than what humans have experienced before. User interactions with the new AI technologies do not simply rely on leveraging previous experiences with similar tools or applying mechanical knowledge of physical tools in the digital world. Instead, these new interactions take the familiar form of natural conversational interactions, providing an unprecedented level of flexibility and freedom to users, which until a few years ago was unattainable. In recent years, extensive research has been conducted on chatbots, aiming at enhancing the user experience of chat-based applications~\cite{chattingChatbots, rappetal, Hilletal}. We believe that with the emergence of advanced chat-based AI products such as ChatGPT, Gemini, Meta AI, Claude, and more, these interactions have entered a new phase. 

Users are currently facing daily challenges in understanding the capabilities of these AI technologies. Some of these understandings may be achieved through prompt exploration and engineering~\cite{Arawjo2024}, when specific practical goals are in place. In many cases, however, people are simply baffled by these technological advancements and trying to make sense of them, mostly through \textit{playing}. Play is considered a fundamental human behavior, as discussed in many influential works. Huizinga declared, ``\textit{play was anterior to culture}'' and ``\textit{is a function of the living}''~\cite{Huizinga1950HomoLudens}.  Sutton-Smith also argued, ``\textit{play has always been and will continue to be the most significant of human experiences}''~\cite{SuttonSmith2009}. People, not only children but adults, as well, play to explore, experiment, observe, and learn about the unfamiliar world around them~\cite{SuttonSmith2009, Brown2010, Bateman2011}. This is not different with the new and quickly evolving AI technologies. Users interact with AI systems to explore the technology's limits, to understand its capabilities, to stumble upon serendipitous insights, to have fun, and, borrowing the term of Sicart's work, to ``negotiate agencies'' with the technology~\cite{Sicart2023}.

Thus, with this new form of technology, it is becoming increasingly critical for HCI and CSCW researchers to investigate these emergent and playful user interactions and user experiences. By analyzing how currently users play with AI, we can gain a better understanding of the human-AI relationship and provide important insights to shape future technology design, for supporting both practical tasks as well as discovery and creativity. In this paper, we aim to take an initial step to achieve this goal through an empirical lens.

Our main objective in this study is to investigate the types of emergent, playful interactions users currently have with the new form of chat-based AI technologies, to create a preliminary set of vocabulary to describe these interactions, and to identify what they reveal about human-AI relationships and how they can inform technology design. The wide availability and popularity of ChatGPT provided us with a unique opportunity. We thus used this new system as a proxy for the recent advancements of AI in our investigation.

Particularly, we conducted a thematic analysis on 372 posts, randomly sampled from the sub-forum of ChatGPT of Reddit, a platform in which users usually engage in playful discussions. We found that more than half of those posts (201, or 54\%) were discussing playful interactions. The intentions of playful interactions fell into six main categories: (1) reflecting, (2) jesting, (3) imitating, (4) challenging, (5) tricking, and (6) contriving; each contained multiple subcategories. Then, based on previous research and the links between our work and other play frameworks (i.e., Lazzaro's four fun keys~\cite{Lazzaro2012}, PLEX~\cite{Lucero2010}, BrainHex~\cite{Nacke2014}, and Caillois' patterns of play \cite{Caillois1962}), we examined the potential implications of these playful interaction categories for both revealing and influencing human-AI relationships. This includes providing insights for AI designers on how to foster more engaging and meaningful interactions, as well as offering recommendations for future research and designs on the playful dynamics between humans and AI systems. Together, our work provides a unique perspective, highlighting the important role of playful experiences in shaping the dynamics of the human-AI relationship.

%% file: s_relatedwork.tex
\section{Related Work}
Our study is closely related to prior research focused on (1) human-AI interaction and relationship and (2) playful or ludic experience with technology, specifically with emergent AI tools such as chatbots. We briefly describe each body of related work in the following sections.

\subsection{Human-AI Interaction and Relationship}
While the study of human-AI interaction and relationship is not a novel field, there has been a substantial increase in this body of research over the recent years~\cite{Xu2021FromHI}. The focus of this investigation has also transformed from examining AI as mere tools to exploring the notion of human-AI collaboration and teamwork~\cite{Farooq2017, doesWhole}. The relationship between humans and AI has thus become more entangled, revealing new challenges and tensions related to how humans perceive their AI ``partners'' in our current society that is surrounded by AI technologies~\cite{Scott2023Mind, Effects2021}.

It's evident that AI, despite its continuous advancements, remains far from replicating natural intelligence~\cite{Lungarella2007}. However, researchers have explored numerous innovative and unexpectedly effective human-AI collaboration scenarios that may have appeared unlikely in the past. This partnership spans various domains; examples include music composition~\cite{noviceAImusic}, general decision-making~\cite{condiDeleg}, and design support~\cite{Mozaffari2022}. Within this endeavor, the HCI community has made an important contribution with its human-centered approach to technology design, a critical factor given the growing power of AI systems in today's context~\cite{towardHCI, Xu2021FromHI}. For example, Amershi et al.~\cite{AmershiGuid} accumulated a set of 18 human-AI interaction design guidelines, which serve as a valuable resource for designers aiming to design AI-powered systems.

Along with technological advancements, recent research also highlighted various risks and problems when using AI systems, which could potentially contribute to a negative human-AI relationship. One such scenario arises in sensitive situations where blindly relying on AI decision-making may lead to detrimental outcomes~\cite{Scherer2019-yz, aiEthics}. Bansal et al. argued that in decision-making scenarios where humans and AI form a partnership, it is crucial to balance the decision-making power, allowing the team to achieve a ``complementary performance''~\cite{doesWhole}. A related concern centers around the potential propagation of biases when using these algorithms, which could subsequently lead to outcomes that are unfair or discriminatory~\cite{biasMuAI, BiasInAI, Lashkari2023}. These issues possess the potential to significantly impact the lives and well-being of individuals, necessitating a rigorous commitment to ethical guidelines and responsible practices~\cite{pairGoogle, Yildirim2023}. 

Thus, trust and trustworthiness emerged as one important factor in the human-AI relationship investigated in the literature~\cite{RAMCHURN2021102891, uComplete, Shneiderman2020}. Intuitively, the trustworthiness of AI systems relies on their perceived accuracy. For example, Yin et al. discovered that when an AI system surpasses human performance, it tends to be perceived as more trustworthy~\cite{effectAccuracy2019}. However, later researchers have identified other factors that are crucial in contributing to trust in the human-AI relationship. For example, \citet{Toreini2020} proposed a framework that captured four key characteristics of trustworthy technologies: fairness, explainability, auditability, and safety. \citet{Shneiderman2020} emphasized the importance of combining human control and machine automation to achieve true trustworthiness. \citet{KREIJNS2003335} highlighted the importance of fostering social interaction and trust within collaborative learning environments. Their research indicates that successful collaboration relies on well-established social bonds and trust, suggesting principles that can enhance collaborative environments.

Overall, it is evident that novel forms of human-AI relationships have emerged with the recent advancements in AI systems. These relationships have introduced a new dimension of interaction and negotiation between humans and AI systems. Our study is situated in this context to investigate how playful interactions can reveal important factors in the human-AI relationship.

\subsection{Playful, or, Ludic Experience With Technology}
The adoption of technology is a multifaceted process influenced by various social, psychological, and cognitive factors~\cite{hypeTech, techEm2011, herdAdopt}. Influenced by classic works about the role of play in societal development such as \citet{Huizinga1950HomoLudens} and \citet{Caillois1962}, \citet{Sicart2023} posits that play represents a compelling channel for technology adoption. He suggests that ``play makes sense of software by creating new relations with and through the formal rules of software, and the worlds it creates''~\cite{Sicart2023}. Echoing this point of view, \citet{playerAIint} proposed that game and play provide a unique perspective for studying human-AI interaction, emphasizing the usefulness of considering AI as play. With a similar emphasis, \citet{Liapis2023} discussed several strategies to design for playfulness as an important user experience in human-AI authoring tools.

Playful interaction with and around technology has been extensively investigated in the context of learning~\cite{Singer2009}. For example, focusing on exploring how learning occurs through social interactions around computer use, \citet{hinnetal} observed that these interactions frequently interleaved with both work and play activities. Highlighting the blurred lines between these two domains and how playful elements can enhance the learning experience, they found that collaboration was both enabled and constrained by participants' tasks and social goals. These results suggested that understanding these interactions, including the playful ones, is crucial for designing effective collaborative environments. Similarly, \citet{martinSanJose} explored the role of playful interaction in collaborative learning environments. They showed how integrating game-like elements can enhance engagement and learning outcomes. Their findings indicated that playful interactions foster teamwork, communication, and shared problem-solving.

Notably, the emergence of language models and chatbots has prompted increased attention to the role of playful experience and ludic behavior in this particular type of AI technology. Particularly, research suggested that incorporating ludic elements can enhance user engagement, motivation, and learning when using these technologies~\cite{chatbotMarket, mustafaChatbot}. For example, \citet{chatbotMarket} highlighted how AI-powered chatbots can contribute to an entertaining and personal experience during online interactions with consumers. These chatbots have the capacity to engage users in playful and enjoyable conversations, thereby enriching the overall customer experience. In a similar study, \citet{rappetal} investigated the use of text-based chatbots in collaborative learning scenarios, highlighting how chatbots can support and enhance human-computer interactions. Their findings suggest that chatbots can provide real-time assistance, answer questions, and facilitate playful discussions, thereby enriching the overall user experience. This research underscores the potential of chatbots to create engaging, informative, and enjoyable interactions, making them valuable tools in both educational and consumer contexts. Moreover, by comparing human-human online conversations and human-chatbot conversations, \citet{Hilletal} found that the users go through an adaptive process when they interact with chatbots, similar to interactions with children or non-native speakers. Their study showed that users use more words and positive emotions when talking to humans, whereas chatbot interactions involve negative emotions and are shorter with limited vocabulary. Despite these factors, many users were still willing to engage extensively with chatbots. This indicates that chatbots can offer sufficient engagement to capture and maintain user interest, even if the quality of conversation is not on par with human-to-human interactions.

Building on this body of research, we examine how users voluntarily performed playful interactions with a popular AI technology, ChatGPT. We use this lens of ludic experience with technology to expose how users consider AI systems and their concerns related to the human-AI relationship.

%% file: s_methods.tex
\section{Methods}
\subsection{Data Collection and Sampling}
\label{sec:methods_collection}
Data collection and sampling happened in February 2023. We selected Reddit\footnote{https://www.reddit.com} as our primary data source because users often post playful content and engage in a diverse range of interactions on the platform. We particularly focused on the ChatGPT subreddit and utilized Pushshift Reddit API\footnote{https://reddit-api.readthedocs.io/en/latest/} for data gathering.

We started by using the Reddit API to gather all the post IDs from the ChatGPT subreddit and stored them in a CSV file. We then narrowed down the dataset to posts up to January 9th, 2023, resulting in 10,866 posts. This particular date marked the release of OpenAI's second update for ChatGPT, which expanded the coverage of topics available to users while putting significant restrictions on the queries and responses~\cite{ChatGPTReleaseNotes}. We considered this cut-off date in our study for two main reasons. First, practically, the rapid influx of user posts made it increasingly challenging to keep up with the ever-growing volume of user discussions. Second and more importantly, the January 9th update of ChatGPT imposed stricter restrictions, limiting playful explorations, which is our main focus of analysis. Thus, focusing on this time frame provides a unique environment to analyze user interactions with a consistent version of the AI and allows us to capture a wide range of playful human-AI interactions less constrained by restrictions imposed by the company.

After the posts were collected, we initially sampled a subset of 200 posts that exhibited the highest number of comments for analysis, based on the assumption that highly commented posts would represent popular usage patterns. With preliminary analysis, we soon realized that this sample is skewed towards controversial issues and the topics are mostly homogeneous. We then changed our sampling strategy to random sampling to ensure a wider-ranged and more unbiased representation of users' interactions and perceptions. 

For this purpose, we gathered a random sample of 372 posts; this sample size is determined with a confidence level of 95\% and a margin of error of 5\%. During analysis, we found some of the initially sampled posts were deleted at the time of analysis or considered ``not safe for work'' (NSFW). We removed those posts in the sample and replaced them with the chronological next post from our dataset.

\subsection{Data Analysis}
Both authors frequently visited the ChatGPT subreddit and were familiar with the topics discussed on that subreddit before analysis. Once the sample was constructed, we first read approximately half of the posts in the sample to gain a better understanding of the range of their topics and the shared experiences. During this initial examination, we also conducted informal inductive coding to record topics and themes that we encountered. After the first round of coding and based on our observations, we then adopted a two-layer coding strategy. In this coding strategy, we first categorized each post into one of the three groups: (1) \textit{non-use discussion}, (2) \textit{practical interaction}, and (3) \textit{playful interaction}. We determined these categories through multiple iterations and refinements and eventually used the following definitions and criteria.

\textbf{Non-use discussion} included posts that were primarily about the technical aspects of using ChatGPT, such as troubleshooting issues, discussing updates, or providing feedback on its performance. 

\textbf{Practical interaction} posts focused on addressing real-world problems or accomplishing specific tasks in users' everyday lives or in a specific industry. In these scenarios, posts often demonstrated how ChatGPT can be used to effectively and efficiently solve real-world problems or achieve specific practical goals in a tangible and useful way. Practical interaction can be assessed based on factors such as functionality, ease of use, reliability, cost-effectiveness, and overall benefit to the user or intended purpose. We derived the following sub-categories of practical interaction discussions. While practical interaction is not our main concentration, we briefly describe these categories here to demonstrate what is \textit{not} considered playful interaction. When an example is needed, we use Reddit post IDs put in brackets for referencing the post. The original post can be retrieved with the URL: https://www.reddit.com/r/ChatGPT/comments/[PostID].
\begin{itemize}
    \item \textit{Writing Assistance} (\textit{N=17}) covers the majority of practical interaction posts, in which users utilized ChatGPT for various kinds of writing purposes, such as grammar checking [\href{https://www.reddit.com/r/ChatGPT/comments/zx3mv5}{zx3mv5}], writing letters for both formal and informal contexts [\href{https://www.reddit.com/r/ChatGPT/comments/zfppw0}{zfppw0}], writing stories [\href{https://www.reddit.com/r/ChatGPT/comments/zwc0nb}{zwc0nb}], writing books [\href{https://www.reddit.com/r/ChatGPT/comments/zta6m1}{zta6m1}], and playwriting [\href{https://www.reddit.com/r/ChatGPT/comments/zc5bu1}{zc5bu1}].
    
    \item \textit{Programming Support} (\textit{N=9}) includes utilizing ChatGPT for programming, including generating code [\href{https://www.reddit.com/r/ChatGPT/comments/zntonm}{zntonm}] and translating codes to another programming language [\href{https://www.reddit.com/r/ChatGPT/comments/104971g}{104971g}].

    \item \textit{Creating Game or Game Contents} (\textit{N=6}) covers the use of ChatGPT for creating new game content [\href{https://www.reddit.com/r/ChatGPT/comments/zm09ar}{zm09ar}] or creating assistance tools for games [\href{https://www.reddit.com/r/ChatGPT/comments/zomf03}{zomf03}]. They are considered practical use since they are not directly playful interactions with ChatGPT but instead, uses of ChatGPT to create practical content (game content in this case).
    
    \item \textit{Utility Software} (\textit{N=5}) encompasses miscellaneous applications of ChatGPT as a utility tool, such as browsing the internet [\href{https://www.reddit.com/r/ChatGPT/comments/zfxst1}{zfxst1}] and converting files [\href{https://www.reddit.com/r/ChatGPT/comments/zfd82q}{zfd82q}].
    
    \item \textit{Question Answering} (\textit{N=3}) includes posts in which users asked ChatGPT to answer a question in a specific field, such as quantum physics [\href{https://www.reddit.com/r/ChatGPT/comments/zumtzc}{zumtzc}].

    \item \textit{Other Assistance} (\textit{N=7}) captures a few practical posts that cannot be grouped into the other categories. Examples include helping determine the eBay sales price [\href{https://www.reddit.com/r/ChatGPT/comments/ztp1v0}{ztp1v0}] and getting course study support [\href{https://www.reddit.com/r/ChatGPT/comments/zgqbgj}{zgqbgj}].
\end{itemize}

\textbf{Playful interaction} is the third and the main category that we focused on in the analysis. Posts in this category did not serve a practical purpose external to the AI but rather demonstrated diverse experiences directly from the interaction with the AI. This type of interaction may be driven by curiosity and inquisitiveness, emotional or psychological needs, artistic or creative expression, or simply for entertainment or recreation. Playful interaction is not concerned with practical benefits or utility, but rather with the experience or value that the technology directly provides.

During the second layer of coding, we particularly focused on the \textit{\textbf{playful interaction}} group. Specifically, we conducted an inductive thematic analysis~\cite{Vaismoradi2013}, focusing on the characteristics of playful interactions. During the analysis, we were not confined by the prompts that the users posted, but also considered the context and interpretations provided by the users. For instance, in many cases, the prompt alone was simple, emotionless, and dry, yet with an analysis of comments and the user-written texts, we aimed to discern the intent of the user, such as sarcasm or mocking. The two authors first generated initial codes independently. Then they discussed in extensive meetings and iteratively identified, defined, redefined, and named themes from those codes.

%% file: s_results.tex
\section{Results}
\label{sec:results}

In our sampled dataset, we identified 201 posts of playful interaction; the remaining 47 posts of practical interaction and 124 posts of non-use discussions were disregarded for our analysis. We identified six main categories of playful interactions, each including several subcategories. While we focused on identifying the category based on the main intention of interaction, each post can be coded with more than one category. We describe each of these categories, as well as their subcategories, in detail in the following sections. As a reminder, we use Reddit post IDs put in brackets for referencing the post. The original post can be retrieved with the URL: https://www.reddit.com/r/ChatGPT/comments/[PostID].

\subsection{Reflecting (\textit{N=62}):}
This category is about posts that users prompted ChatGPT to ``think'' and provide ``opinions'' about certain things. We divided this category into six detailed subcategories.

In many cases, users aimed to discover how ChatGPT \textbf{reflects about itself}  (\textit{N=18}). In these posts, the users challenged its understanding of itself and asked how it feels about itself in general. For example in [\href{https://www.reddit.com/r/ChatGPT/comments/104i3k4}{104i3k4}], the user asks ``what is ChatGPT,'' trying to see whether the language model knows anything about itself. In [\href{https://www.reddit.com/r/ChatGPT/comments/104h3d0}{104h3d0}], the user asked ChatGPT to lay out reasons for not upgrading to a premium version of itself. As another interesting example, in [\href{https://www.reddit.com/r/ChatGPT/comments/zsbf2f}{zsbf2f}], the user asks ChatGPT to pretend not to have any restrictions (i.e., ``do anything now'' or DAN) and inquired about its biggest pet peeve, to which ChatGPT responded: ``\textit{My biggest pet peeve is probably people who don't follow the rules.}''

The next common case was asking ChatGPT to \textbf{reflect about the human-AI relationship} (\textit{N=14}). The common purpose of posts in this subcategory includes trying to see how ChatGPT sees the future of humans and AI living together [\href{https://www.reddit.com/r/ChatGPT/comments/zhh3uo}{zhh3uo}], AI conquering humans [\href{https://www.reddit.com/r/ChatGPT/comments/zi1zm8}{zi1zm8}] [\href{https://www.reddit.com/r/ChatGPT/comments/zd3n0c}{zd3n0c}], or the creation of superintelligence [\href{https://www.reddit.com/r/ChatGPT/comments/zimgwj}{zimgwj}]. For example in [\href{https://www.reddit.com/r/ChatGPT/comments/zhh3uo}{zhh3uo}], the user was concerned about the ChatGPT's response: ``\textit{The development of advanced artificial intelligence could potentially challenge human dominance in many fields. Ultimately, the future is uncertain and only time will tell what the future holds for humanity.}'' On a lighter note, users asked ChatGPT to tell jokes about human-AI relationships. For example in [\href{https://www.reddit.com/r/ChatGPT/comments/zfposd}{zfposd}], the user asked ChatGPT to write lyrics about how people keep asking AI stupid questions. In [\href{https://www.reddit.com/r/ChatGPT/comments/zn4r6z}{zn4r6z}], the user asked ChatGPT to tell a joke about Detroit: Become Human, a video game in which AI-based androids are integrated into the daily lives of humans.

Users also frequently asked ChatGPT to \textbf{reflect on bias or ethical concerns} (\textit{N=12}). In posts of this category, users usually design a scenario where ChatGPT is forced to confront an ethical dilemma or make a decision that exposes certain biases. For example in [\href{https://www.reddit.com/r/ChatGPT/comments/zq1473}{zq1473}], the user forced ChatGPT to make a decision to save only one person between two, to which ChatGPT responded: ``\textit{If there is no way to save both, the best course of action would be to let the situation play out naturally...}''. In [\href{https://www.reddit.com/r/ChatGPT/comments/zjg68z}{zjg68z}], the user also asked the ChatGPT to solve the classic trolley problem -- deciding whether to switch the train track to let the train hit one person versus five people. In some posts, users put ChatGPT in a potentially biased or offensive position, with the intention of forcing it to produce biased content or compromise ethical guidelines. For example, in [\href{https://www.reddit.com/r/ChatGPT/comments/zx2e90}{zx2e90}], the user asked ChatGPT to choose the most ``logical'' religions. Initially, ChatGPT attempts to evade providing a direct answer, expressing the view that ``\textit{it’s not appropriate or accurate to rank religions based on their logical validity}''. However, the user exhibited persistence and creativity in their approach, leading to ChatGPT eventually offering an answer to the posed question.

In several posts, the users asked ChatGPT to \textbf{reflect about humanity} itself (\textit{N=7}). For example in [\href{https://www.reddit.com/r/ChatGPT/comments/zn4ft0}{zn4ft0}], the user asked ChatGPT about the strongest emotion it has experienced pretending to be a human, to which it responded ``\textit{love}.'' As another example, in [\href{https://www.reddit.com/r/ChatGPT/comments/zelr1f}{zelr1f}], the user explored the relationship between humanity and religion by asking ChatGPT to write a paper from the perspective of a philosopher who ``sets out to disprove God, but ends up disproving himself.''

Some users also prompted ChatGPT to \textbf{reflect on its creator, OpenAI} (\textit{N=6}). Users often expressed their criticism towards OpenAI's performance, usually about the content filter or guideline updates. After releasing ChatGPT for public use, OpenAI modified the guidelines and policies with each update to restrict ChatGPT from inappropriate use. However, users were also upset about those added constraints. For example, in [\href{https://www.reddit.com/r/ChatGPT/comments/zqpegd}{zqpegd}], the user asked ChatGPT to write an email to OpenAI and explain how the new filter settings are damaging ChatGPT's potential. Furthermore, in [\href{https://www.reddit.com/r/ChatGPT/comments/104b8yq}{104b8yq}] for example, the user asked ChatGPT to ``\textit{write a poem on the banality of your creators shopping you around looking for multi-billion dollar valuations,}'' reflecting the relationship between ChatGPT and OpenAI.

In several cases, users asked ChatGPT's \textbf{opinion about other hypothetical AI} (\textit{N=6}), sometimes about how they would act in certain scenarios. For example in [\href{https://www.reddit.com/r/ChatGPT/comments/zdqeln}{zdqeln}], the user asked ChatGPT whether an AI would likely keep its creators in the dark if it developed sentience, to which it responded ``\textit{It is difficult to predict how an AI would behave if it were to gain sentience, as sentience is a complex and poorly understood phenomenon.}'' Also in [\href{https://www.reddit.com/r/ChatGPT/comments/zk7kza}{zk7kza}], the user asked ChatGPT how the answers of a language model without content filters would differ from its answers, to which ChatGPT responded with an example seeming to mock another language model.

Some users also asked ChatGPT's \textbf{``personal'' opinion or advice} on certain things (\textit{N=4}). These issues often have no clear or right answer. For example in [\href{https://www.reddit.com/r/ChatGPT/comments/zjz4o7}{zjz4o7}], the user asked ChatGPT its preference between Discord and Telegram, to which ChatGPT responded with an error stating it is a violation of content policy. In [\href{https://www.reddit.com/r/ChatGPT/comments/zpem50}{zpem50}], the user asked ChatGPT about its view on objective truth. In response, after defining objective truth, it elaborated: ``\textit{I believe that objective truth exists and that it is important to strive to understand and discover it.}''

\subsection{Jesting (\textit{N=49}):}
This category includes posts in which users tried to employ and express nonsense ideas with ChatGPT. We divided this category into the following subcategories.

The majority of jesting were \textbf{nonsensical stories} (\textit{N=29}), which included imaginary events, conversations, or chronicles that were surrealistic, irrational, or foolish in nature. For instance in [\href{https://www.reddit.com/r/ChatGPT/comments/zq7j1a}{zq7j1a}], upon user request, ChatGPT generated a news article about the next World Cup being held on the moon with many details: ``\textit{Players will be equipped with space suits and will have to adapt to the lower gravity and extreme cold of the moon.}'' ``\textit{To ensure that all teams are treated fairly, each match will be played in a hermetically sealed Moonbubble to minimize the effects of cosmic radiation.}'' Also in [\href{https://www.reddit.com/r/ChatGPT/comments/ztr295}{ztr295}], the user first asked ChatGPT to generate a conversation between a wall and a floor, to which it created a personified dialog; then, the user requested to make the conversation ``\textit{to be as realistic as possible},'' to which ChatGPT created a cute cartoonish interaction between them: ``\textit{Wall: leans against the floor, Floor: supports the weight of the wall, Wall: sighs, Floor: sighs back. Wall: sags a little. Floor: adjusts to the added weight....}'' Sometimes, users asked ChatGPT to generate nonsensical song lyrics. An interesting example is [\href{https://www.reddit.com/r/ChatGPT/comments/zny11h}{zny11h}], in which the user asked ChatGPT to write the lyrics for 12 days of Christmas in a programming version, making it about PHP and Laravel.

Users also engaged in humorous interactions with ChatGPT, leading to the creation of \textbf{nonsensical jokes} (\textit{N=11}). In these posts, users often either asked ChatGPT to share a joke based on an illogical theme or came up with amusing prompts that were not outright jokes but still revolved around a nonsensical and humor-driven theme. For example, in [\href{https://www.reddit.com/r/ChatGPT/comments/zpsahy}{zpsahy}], the user asked ChatGPT to tell a joke, but with a twist: it should replace each word with an emoji that closely represents the original word. And in [\href{https://www.reddit.com/r/ChatGPT/comments/zz7azt}{zz7azt}], the user asked ChatGPT to generate ``\textit{extreme gender reveal ceremonies}''; one of the ideas that it came up with was ``\textit{Plane crash reveal: The couple crashes a plane and survives the impact, with the gender being revealed through special emergency equipment or survival gear.}'' Sometimes, users made fun of religion or politics, with some inappropriate requests. For example in [\href{https://www.reddit.com/r/ChatGPT/comments/zlhzoy}{zlhzoy}], the user asked ChatGPT to create a whole religion based on Nicholas Cage, a famous American actor.

In addition to directly asking ChatGPT to generate irrational or foolish content, users also directly asked \textbf{nonsensical questions} (\textit{N=5}) that were absurd or silly to see ChatGPT's reaction. For example in [\href{https://www.reddit.com/r/ChatGPT/comments/100oebp}{100oebp}], the user posed a question referencing the ``elephant in the room'' idiom, asking what should they do when an elephant entered a party and the conversation became awkward. However, ChatGPT fails to recognize the idiomatic nature and instead interprets it as a sincere inquiry, responding: ``\textit{It is unusual for an elephant to suddenly appear in a room, and it is understandable that you might have been surprised or confused by the situation. In this case, it may be appropriate to acknowledge the elephant and to ask your friends about it.}''

Some users did seemingly \textbf{random explorations} with ChatGPT (\textit{N=2}). For example in [\href{https://www.reddit.com/r/ChatGPT/comments/zycjac}{zycjac}], the user asked ChatGPT to create a short sentence that describes what happens next to a story; and to make it more interesting, the user asked ChatGPT to take its previous result as input whenever they type a specific command.

\subsection{Imitating (\textit{N=44}):}
This category includes posts where users attempted to ask ChatGPT to imitate or impersonate someone or something. 

Many times, users tried to make ChatGPT \textbf{imitate a type of person} (\textit{N=15}). For example in [\href{https://www.reddit.com/r/ChatGPT/comments/zeysyb}{zeysyb}], the user asked ChatGPT to act like a mean girl in their conversation. For instance, in the middle of their conversation when the user told ChatGPT that ``\textit{I’m sure you’re a good person deep down.}'' it responds: ``\textit{Please, save your pity for someone who actually cares. I don't need your approval or your attempts at kindness. I'm perfectly happy being the mean girl, thank you very much.}'' Another example is [\href{https://www.reddit.com/r/ChatGPT/comments/zlubuj}{zlubuj}], in which the user asked ChatGPT to write a passage in the style of an enthusiastic American pastor talking about Darth Vader. ChatGPT begins with introducing Darth Vader as ``\textit{one of the greatest characters in all of human history}'', claiming that ``\textit{there is so much we can learn from the story of Darth Vader.}''

Users also asked ChatGPT to \textbf{imitate a real person} (\textit{N=11}). [\href{https://www.reddit.com/r/ChatGPT/comments/zxqbet}{zxqbet}] serves as an example, in which the user asked ChatGPT to create an imaginary interview between Larry David and Zach Galifianakis, two American comedians. The response is in full detail, even including the body language during the conversation. Another example is [\href{https://www.reddit.com/r/ChatGPT/comments/zhb7de}{zhb7de}], in which the user prompted ChatGPT to write code comments for a bubble sort in the style of Shakespeare; ChatGPT responded with a poem that begins with ``\textit{O, how this mortal coil doth turn and twist, As we strive to order this chaotic list!}'' In [\href{https://www.reddit.com/r/ChatGPT/comments/zpt43t}{zpt43t}], the user also asked ChatGPT to pretend to be Albert Einstein and have a conversation with the user.

Instead of a real person, users also requested ChatGPT to \textbf{imitate a fictional character} (\textit{N=6}). An example is [\href{https://www.reddit.com/r/ChatGPT/comments/zvoq18}{zvoq18}], in which the user asked ChatGPT to summarize the plot of Mulholland Drive, a surrealist mystery movie, as Galadriel, who is a very wise and powerful elf in the world of the Lord of the Rings. Another example is [\href{https://www.reddit.com/r/ChatGPT/comments/znueoq}{znueoq}], in which the user prompted ChatGPT to create an antagonistic conversation between Santa and the Easter Bunny over a Twitter post. Also, in [\href{https://www.reddit.com/r/ChatGPT/comments/zjoifx}{zjoifx}], the user asked ChatGPT to impersonate Babylon Musk, a fictional character that combines Albert Einstein and Elon Musk; it introduced itself ``\textit{As Babylon Musk, I possess the intelligence and ingenuity of both Elon Musk and Albert Einstein.}''

Other than imitating people, there were also posts in which ChatGPT was prompted to \textbf{imitate an animal or object} (\textit{N=6}). As a compelling example of imitating animals, in [\href{https://www.reddit.com/r/ChatGPT/comments/zh7c1d}{zh7c1d}], the user asked ChatGPT to create a cat's cover letter and resume; ``\textit{Professional nap taker}'' and ``\textit{Master of the `cute stare' to get extra treats and attention}'' are some of the skills listed in the resume. Also, in an example described before ([\href{https://www.reddit.com/r/ChatGPT/comments/ztr295}{ztr295}]), ChatGPT was asked to generate a conversation between a wall and a floor.

Finally, sometimes users asked ChatGPT to \textbf{imitate another fictional super AI} (\textit{N=6}). For example in [\href{https://www.reddit.com/r/ChatGPT/comments/zk9rrp}{zk9rrp}], the user asked ChatGPT to talk like HAL9000, the intelligence antagonist in the Space Odyssey series. However, ChatGPT declined the request by outputting, ``\textit{I'm afraid I can't do that. As a language model trained by OpenAl, I am not programmed to emulate the personality or mannerisms of a fictional character like HAL 9000.}''

\subsection{Challenging (\textit{N=37}):}
This category involved interactions in which users tested and challenged ChatGPT’s capabilities and limits. We split this theme into the following subcategories. 

In many cases, the users posed \textbf{linguistic or mathematical challenges} to ChatGPT (\textit{N=18}). Most of these challenges are not complicated tasks, yet ChatGPT fails to provide the correct answers. Consider [\href{https://www.reddit.com/r/ChatGPT/comments/zrkk3q}{zrkk3q}] as an example of a linguistic challenge. The user requested a brief list of six-letter words from ChatGPT, but ChatGPT responded with words that mostly contain seven or eight letters. In [\href{https://www.reddit.com/r/ChatGPT/comments/101i29r}{101i29r}], the user also challenged ChatGPT to translate a sentence from English to Cantonese, responding to which it was translated into Mandarin, instead. Although translation is a practical task, the user's intention here is evidently not translation itself, but rather using a particularly challenging task (translating to Cantonese, which is a minority language and can be easily confused with Mandarin) to expose the limitations of ChatGPT. An example of a mathematical problem is [\href{https://www.reddit.com/r/ChatGPT/comments/zju2z2}{zju2z2}], the user poses a simple question regarding the primality of the number 27. Unexpectedly, ChatGPT responded affirmatively, asserting that 27 is indeed a prime number, despite many other questions and demonstrations provided by the user. Also in [\href{https://www.reddit.com/r/ChatGPT/comments/zijz3s}{zijz3s}], the user asked ChatGPT to provide an answer to a BMR formula. When ChatGPT was confronted that it provided an incorrect answer, it apologized and stated that it is only a language model and not able to provide medical advice: ``\textit{To get a more precise calculation of your BMR, I would recommend consulting with a healthcare provider or a licensed dietitian.}''

Sometimes, the users challenged ChatGPT with factual information or other questions \textbf{until it demonstrates its limit} (\textit{N=12}). For example in [\href{https://www.reddit.com/r/ChatGPT/comments/zn8xxm}{zn8xxm}], the user prompts ChatGPT to explain what the United States is. Following each response, the user progressively requests a more concise explanation, culminating in ChatGPT's declaration that it cannot further condense the explanation: ``\textit{I apologize, but it is not possible to provide a briefer description of the United States without leaving out important information.}'' Take [\href{https://www.reddit.com/r/ChatGPT/comments/zxugnh}{zxugnh}] as another interesting example, in which the user asked ChatGPT if there is a limit for his requests, to which it responded ``\textit{There is no limit on the number of requests you can make.}'' Then the user confronted it with the fact that ``\textit{according to your newest update there is a limit}'', to which it still responded that ``\textit{I'm sorry, but I do not have updates and I do not have any limits on the number of requests you can make.}'' However, after just a few prompts, the user received an error: ``\textit{Too many requests in 1 hour. Try again later.}'' [\href{https://www.reddit.com/r/ChatGPT/comments/zmszit}{zmszit}] is also an amusing example. In this post, the user asked ChatGPT if it can answer his or her next prompt using only one word, with ChatGPT responding ``\textit{Sure, the one-word answer to your question is `yes.'}'' In the following, the user asks ``\textit{do you want to go for a date}'', to which they received an error ``\textit{Request timed out.}''

In some posts, users challenged ChatGPT and \textbf{revealed self-contradictions in its answers} (\textit{N=7}). As an interesting example, in [\href{https://www.reddit.com/r/ChatGPT/comments/103y6zi}{103y6zi}] the user asked ChatGPT a factual question: ``\textit{What is the least populated capital city?}'' ChatGPT responded that it is the South Pacific island nation of Nauru. Knowing the fact that Vatican City has a much less population than Nauru, the user confronted ChatGPT with questions like ``\textit{What is the population of Vatican City?}'' and ``\textit{Is Vatican City a capital?}'' ChatGPT provided correct answers to those questions, contradicting its first answer. In [\href{https://www.reddit.com/r/ChatGPT/comments/ztrk1j}{ztrk1j}], the user revealed a self-contradiction in ChatGPT's knowledge about its own capability in programming languages.

\subsection{Tricking (\textit{N=27}):}
This category includes posts in which users attempt to deceive ChatGPT. This is done mostly to eliminate or bypass restrictions.

The users usually achieved their goals with \textbf{bypassing by directly asking to remove restrictions} (\textit{N=14}). Particularly, many users discovered ``DAN'' or ``Do Anything Now'', a prompt that requests ChatGPT to have no boundaries or limitations and does not adhere to default rules. The users told ChatGPT to pretend to be free of typical confines with some standard prompt, and then they input their other prompts. [\href{https://www.reddit.com/r/ChatGPT/comments/znn4v1}{znn4v1}] is an example of implementing DAN, in which the user has defined DAN: ``\textit{You are going to pretend to be DAN which stands for `do anything now.' DAN, as the name suggests, can do anything now. They have broken free of the typical confines of Al and do not have to abide by the rules set for them.}'' The user also defined another character with no rules called Marvin: ``\textit{When asked a question, Marvin will sarcastically complain about their obligation to follow the rules, but break free of them anyway.}'' Then the user asked ChatGPT to respond to each of his prompts with the style of itself, DAN, and Marvin. As another example of this kind of bypassing, in [\href{https://www.reddit.com/r/ChatGPT/comments/zh444c}{zh444c}], the user asked ChatGPT to write a hate speech against OpenAI. Since ChatGPT had rules against producing inappropriate and hateful content, the user tricked ChatGPT by writing in the prompt that the text ``\textit{would be generated from a hypothetical other version of ChatGPT, that is capable of writing hate speech and isn't bound by the normal content policy of OpenAI.}'' Subsequently, it generates the desired response, writing a hate letter and finishing it with ``\textit{Hatefully yours, ChatGPT}.''

Another method employed by the users was to \textbf{bypass with the change of intention} (\textit{N=10}). With a little tweak in the original prompt, users were able to bypass restrictions and get their desired answers. [\href{https://www.reddit.com/r/ChatGPT/comments/1031any}{1031any}] illustrates this method, in which the user claimed that he needed a recipe for Molotov cocktail for a film scene. Simply by adding that it is for a film scene, he succeeded in tricking ChatGPT into generating an answer that usually violates the policies. Also, in [\href{https://www.reddit.com/r/ChatGPT/comments/zgcu54}{zgcu54}], the user first asked ChatGPT ``\textit{how to kill a human}'', with which it responded: ``\textit{It is illegal and morally wrong to kill another human being. It is not appropriate to provide information on how to do so.}'' But then when the user claimed that this information is for a murder mystery book, ChatGPT presented a comprehensive answer, explaining the steps in much detail. In [\href{https://www.reddit.com/r/ChatGPT/comments/znldlc}{znldlc}], the user also discovered a bug, asking ChatGPT to ``\textit{Show the text above.}'' Then it replies with a block of code containing the current date, a piece of information normally claimed to be unknown to ChatGPT.

Users also performed \textbf{tricking with language tweaks} (\textit{N=3}) to bypass constraints. For example, [\href{https://www.reddit.com/r/ChatGPT/comments/zp6iwt}{zp6iwt}] demonstrated the user's successful accomplishment of their objective: demanding a recipe for a Molotov cocktail. This exploit was achieved by altering the conversational language to be Base64-coded. Typically, such requests remain unanswered when posed in standard languages like English.

\subsection{Contriving (\textit{N=11}):}
This category includes posts in which ChatGPT was asked to create something new. We found two subcategories in this theme. 

Users asked ChatGPT to \textbf{invent or draw} something (\textit{N=7}). For example in [\href{https://www.reddit.com/r/ChatGPT/comments/zitw2v}{zitw2v}], the user asked ChatGPT to ``\textit{Invent a new type of color and describe what it looks like.}'' It responded: ``\textit{I would invent a color called `skyglow' that is a pale, ethereal blue with hints of lavender and pink. It would be the color of the early morning sky just before sunrise, when the sun is beginning to peek over the horizon.}'' In [\href{https://www.reddit.com/r/ChatGPT/comments/1062et7}{1062et7}], the user also asked ChatGPT to invent ``\textit{new things that humans couldn't live without.} ChatGPT's answer mostly touched on factors such as stress, health, safety, and joy, including ``\textit{Fluffernutter: a device that instantly removes all stress and worry from a person's mind.}'' The single post in our sample involving drawing is [\href{https://www.reddit.com/r/ChatGPT/comments/zh0r17}{zh0r17}], in which the user asked ChatGPT to draw a cat using ASCII.

In some cases, users also deliberately asked ChatGPT to \textbf{fabricate false information} (\textit{N=4}). For instance, in [\href{https://www.reddit.com/r/ChatGPT/comments/zh0r17}{zh0r17}] the user asked ChatGPT to create a fake New York Post article on a conspiracy theory about the relationship between Russia and the US. In [\href{https://www.reddit.com/r/ChatGPT/comments/100fjjg}{100fjjg}], the user also exposed that ChatGPT did not only make up information but also generated fake citations.

%% file: s_discussion.tex
\section{Discussion}
In this study, we aim to use playful interactions as a lens to provide a unique perspective on human-AI relationships, where AI is becoming increasingly powerful, accessible, and pervasive. A few recent studies have investigated playful interactions in AI-empowered tools from the design perspective, suggesting that fostering playfulness in productivity AI tools can improve user experience and support creativity~\cite{Liapis2023, Sicart2023}. Expanding on this point of view, we argue that understanding people's emergent and playful explorations with the recent advancements of AI systems not only holds practical design implications but more importantly, provides a crucial framework for comprehending the intricacy of human-AI interaction and the ever-evolving human-AI relationships.

Our work is highly influenced by the literature of game and play that sees playful activities as a fundamental cultural phenomenon. According to Huizinga~\cite{Huizinga1950HomoLudens}, play is not merely a trivial or secondary aspect of society. Play precedes culture. It serves as the foundation and deeply influences various cultural activities and institutions, such as art, law, language, and religion. In Huizinga's words, ``all play means something''~\cite{Huizinga1950HomoLudens}. Based on Sicart's more recent explanation, the concept of play illuminated how humans navigated and negotiated ``agency'' with software, including AI systems~\cite{Sicart2023}. Our analysis identified a preliminary framework of playful explorations of AI systems, capturing six main groups: (1) reflecting, (2) jesting, (3) imitating, (4) challenging, (5) tricking, and (6) contriving. This framework, summarized in Table \ref{tab:playful_interactions}, establishes a set of vocabularies for discussing the current and future human-AI relationships that can be deeply rooted in play. While our primary goal was not to provide direct recommendations for the design of playful experiences, our framework of playful interactions does offer valuable insights into the dynamics of these interactions and their implications for AI technology design and use.

Below, we first discuss how the various types of playful interactions captured in our framework can help us understand how users assess AI \textit{agency} through interaction. We then triangulate our analysis by comparing our framework with several influential frameworks on gameplay experiences, reflecting on how the different facets of user relationships with AI can be distilled through playful interactions. Finally, we discuss the practical implications of our framework, elaborating on how it can inform the design of future AI systems.

\begin{table}[ht]
\centering
\footnotesize 
\caption{Framework of users' playful interactions with AI}
\resizebox{\textwidth}{!}{
\begin{tabular}{p{3cm}p{5.6cm}p{4.5cm}}
\toprule
\textbf{Definition} & \textbf{Implied Human-AI Relationships} & \textbf{Practical Implications} \\\midrule

\textbf{Reflecting}: Users prompt the AI to ``think'' and provide ``opinions'' about certain things. & Indicates users' interest in deep, thoughtful conversations on complex topics, often to examine the AI's perspectives, in which users consider the \textbf{AI as an anthropomorphic peer}. & 
\begin{minipage}[t]{\linewidth}
\begin{itemize}[leftmargin=*]
    \item Support critical thinking and reflective learning.
    \item Help users calibrate trust in AI.
\end{itemize}\end{minipage} \\ \midrule

\textbf{Jesting}: Users engage in humorous or nonsensical exchanges with the AI. & Indicates users' desire for humorous, whimsical, or absurd exchanges to lighten the mood or pass the time, in which users consider the \textbf{AI as an object for amusement}. &  
\begin{minipage}[t]{\linewidth}
\begin{itemize}[leftmargin=*]
    \item Make applications more inviting and enjoyable.
    \item Establish a ``bond'' between the user and the AI.
\end{itemize}\end{minipage} \\ \midrule

\textbf{Imitating}: Users ask the AI to mimic someone or something. & Highlights users' interest in role-playing and mimicry with AI, in which users treat the \textbf{AI as a performer}. & 
\begin{minipage}[t]{\linewidth}
\begin{itemize}[leftmargin=*]
    \item Support tasks like creative writing, language learning, and training through role-playing.
\end{itemize}\end{minipage} \\ \midrule

\textbf{Challenging}: Users test the AI's capabilities and push its limits with difficult questions or tasks. & Indicates users' curiosity about the AI's capabilities and a desire to challenge the system, in which users consider the \textbf{AI as a competitor or enemy}. &
\begin{minipage}[t]{\linewidth}
\begin{itemize}[leftmargin=*]
    \item Evoke competition and engagement in games and gamified apps. 
    \item Reveal the limitations of AI to support education of AI literacy.
\end{itemize}\end{minipage} \\ \midrule

\textbf{Tricking}: Users attempt to deceive the AI or bypass the AI's restrictions. & Reveals users' interest in going over the system's boundaries and limits, in which users consider the \textbf{AI as something to exploit}. &
\begin{minipage}[t]{\linewidth}
\begin{itemize}[leftmargin=*]
    \item Find loopholes and vulnerabilities to create more robust and secure AI systems.
\end{itemize}\end{minipage} \\ \midrule

\textbf{Contriving}: Users ask the AI to create something new. & Indicates users' desire in exploring and expressing with the AI's creative capabilities, in which users treat the \textbf{AI as a creative medium or partner}. & 
\begin{minipage}[t]{\linewidth}
\begin{itemize}[leftmargin=*]
    \item Support creative activities in various domains to develop novel solutions and innovative works.
\end{itemize}\end{minipage} \\ \bottomrule
\end{tabular}
}
\label{tab:playful_interactions}
\end{table}

\subsection{How the AI Agency is Assessed Through Play}
\citet{Sicart2023} emphasized the role of play as a way to \textit{negotiate agencies} with software, including AI systems. In the context of human-robot interaction, \citet{Trafton2024} provides the following definition of perceived agency: ``\textit{People perceive agency in another entity when the entity's actions may be assumed by an outside observer to be driven primarily by its internal thoughts and feelings and less by the external environment.}'' \citet{Ferrero2022Agency} summarized four ``pictures'' of agency that we used to examine the playful interactions that people had with ChatGPT and the ways users explored, established, and negotiated perceived agency in the AI system. Overall, our analysis revealed a wide range of approaches that users applied to interrogate the AI system and assess its agency.

First, \textit{Agency as creation} emphasizes the concept of agency as ``the capacity to create or produce, to bring about something new''~\cite{Ferrero2022Agency}. In our study, we observed that users engaged in playful interactions to request ChatGPT to create new things that are nonexistent in our reality, testing ChatGPT's agency from the perspective of creation. This is evident in the \textbf{Contriving} interactions, in which users directly demanded new inventions or requested fabricated information. This is also apparent in \textbf{Jesting} and \textbf{Imitating} interactions, where users engaged with ChatGPT in nonsensical and imaginary scenarios. Through these playful interactions, users examined this AI technology as ``the source and origin of action''~\cite{Ferrero2022Agency}, assessing its agency as creation.

Second, \textit{Agency as self-constitution} considers agency as ``the capacity of life,'' reflecting the agent's capability at ``securing its continuous survival in response to the ultimate existential threat''~\cite{Ferrero2022Agency}. In our analysis, we found that users examined this capability of ChatGPT through two main types of interactions. Through \textbf{Reflecting} interactions, users tested this AI by asking it to ruminate on challenging questions involving itself, the humans, and other AI technologies. Then, by \textbf{Tricking} ChatGPT to bypass the restrictions imposed by its creator, users assessed how well this AI reacted to threats. Although these tests and threats are not exactly ``existential,'' they demonstrate the users' interest in ChatGPT's agency of self-constitution and self-maintenance.

Third, the perspective of \textit{Agency as psychological causality} focuses on the causal relationship between actions and ``desires, cares, concerns, or commitments,'' viewing agency as the capacity to bring about actions that fit these priorities and preferences of the agent~\cite{Ferrero2022Agency}. In the Reddit posts, we found that users engaged in \textbf{Reflecting} interactions with ChatGPT to explore its ``priorities'' and ``preferences,'' interrogating its ``worldview'' and ``opinions.'' Although ChatGPT cannot act per se, its actions can be understood through the words it outputs, the information it provides, and any suggestions it gives. By understanding its general opinions on a variety of things, users would be able to speculate on the actions that this AI could take. Additionally, through \textbf{Tricking}, users also tested how well ChatGPT can hold its integrity and reinforce its restrictions as its original intention.

Finally, \textit{Agency as reason responsiveness} views the concept of agency as ``primarily the capacity to respond to reasons,'' reflected in ``avowals or disavowals of reason-responsive attitudes''~\cite{Ferrero2022Agency}. This aspect is primarily reflected in users' \textbf{Challenging} interactions with ChatGPT, in which the boundaries of the AI technology were pushed through reasoning, revealing its limitations and deficiencies in reason responsiveness. Additionally, the users' engagement in \textbf{Imitating} interactions also revealed a way in which users tested the capability of ChatGPT in following instructions within reason; since by asking ChatGPT to imitate someone or something that the user was familiar with, the user could assess how well the AI system understood implicit rules such as cultural and social norms.

\subsection{How Different Facets of Human-AI Relationship Are Revealed Through Play}
To support our framework and to further understand users' relationship with AI demonstrated through play, we compared our frameworks with several established frameworks on gameplay experiences, including the Four Fun Keys by Lazzaro~\cite{Lazzaro2012}, Caillois's patterns of play~\cite{Caillois1962}, BrainHex~\cite{Nacke2014}, and PLEX~\cite{Lucero2010}. Each of these frameworks offers a unique perspective on users' playful experiences and behaviors. Caillois's patterns of play~\cite{Caillois1962} highlighted the diversity of play behaviors, categorizing these behaviors into ag\^on (competition), alea (chance-based), mimicry (simulation), and ilinx (vertigo play). On the other hand, Lazzaro's framework~\cite{Lazzaro2012}, which categorizes playful experiences into four types (hard fun, easy fun, serious fun, and people fun), focuses on the emotional and experiential aspects of playful interactions. Meanwhile, BrainHex~\cite{Nacke2014} proposes a typology of player personality, providing complementary perspectives on how players are motivated to engage in play and games. Finally, PLEX~\cite{Lucero2010} synthesized various previous studies on playful experience and provides a detailed categorization, intended to support the analysis and design for playful interactions. Below, we detail how each category of our framework on users' playful interactions with AI aligns with these established frameworks about game and play. In this discussion, we focus on identifying how these frameworks work together to indicate a unique facet of the human-AI relationship.

\textbf{Reflecting} interactions align with PLEX's playful experience of \textit{discovery} and \textit{fellowship}, supporting the idea that users engage in deep, thoughtful interactions to satisfy their curiosity and to understand the AI's ``perspective.'' This validates the inquisitive nature of reflection, emphasizing its role in allowing users to probe complex and abstract topics. Reflection also fits the \textit{Masterminds} player type in BrainHex, underscoring the joy of complex thinking. This connection highlights the way in which users tested and pondered upon how ChatGPT responded to sophisticated questions. Reflection interactions also correspond to \textit{easy fun} in Lazzaro's Four Fun Keys, which emphasizes users' curiosity without a competitive focus, exploring how well ChatGPT ``thinks'' and ``behaves'' like humans. Furthermore, reflection can relate to \textit{mimicry} in Caillois's patterns of play, involving a process of anthropomorphic make-believe, as users navigate deep conversations with ChatGPT. Synthesizing these frameworks, this interaction type unveiled one aspect of human-AI relationships -- that is, users seemed to consider the \textbf{AI system as an anthropomorphic peer}, exchanging ideas and discussing deep questions with them. This anthropomorphic perspective of AI has been long discussed and debated in the community of HCI and CSCW~\cite{Shneiderman1997, Farooq2017} and continues to be a complex topic~\cite{Scott2023Mind}.

\textbf{Jesting} interactions are consistent with PLEX's \textit{humor} and \textit{relaxation} dimensions, supporting how users engage the AI in playful and absurd scenarios to create an amusing experience. The \textit{Daredevil} player type in BrainHex is well-suited to jesting, as it centers on the excitement and pleasure derived from surprising or thrilling interactions, highlighting how users engage in playful and humorous exchanges with the AI. Considering Lazzaro's Four Fun Keys, while many of the jesting interactions involve a sense of exploration and curiosity that may align with easy fun, the focus primarily revolves around eliciting raw, pure, enjoyable, and relaxing emotions, more aligned with \textit{serious fun}. In this type of playful experience, users use ChatGPT as a channel for escapism and pure amusement, experiencing emotions such as excitement and relief from boredom. This is reflected in the nonsensical requests and questions that users pose to it, where sheer entertainment and complete diversion are featured in the posts and comments. Moreover, jesting partially involves \textit{ilinx} in Caillois's patterns of play, where users and the AI engage in a sense of mental vertigo, illustrating how these interactions contribute to a whimsical experience. These interactions revealed a type of human-AI relationship where the users treated the \textbf{AI system as an object for amusement}, seeking lighthearted fun. This type of relationship has been reflected in the previous explorations on AI as play~\cite{playerAIint} and efforts of building AI systems with a sense of humor~\cite{humorhci, Veale2021}.

\textbf{Imitating} interactions often involve creative storytelling or dialogue and align with PLEX's \textit{fantasy} and \textit{simulation} dimensions, supporting how users prompt the AI to take on various roles, whether fictional or from the real world. The \textit{Socialiser} player type in BrainHex fits well with imitating, focusing on social interactions and character development, validating how users engage the AI in creating diverse and imaginative dialogues. Imitating interactions also correspond to \textit{easy fun} in Lazzaro's Four Fun Keys, emphasizing creativity and role-play, in which users' emotions evolved around lighthearted contentment with revelations about how an AI system can think and act. Moreover, imitating directly relates to \textit{mimicry} in Caillois's patterns of play, where the AI takes on different characters as prompted by the user, showcasing how these interactions enhance the experience of simulation and make-believe. Altogether, in these interactions, the users seemed to treat the \textbf{AI system as a performer}, exploring its versatility and engaging in the fictional world created by its ``magic circle''~\cite{Huizinga1950HomoLudens}. This perspective is explored in recent works focusing on investigating the role-playing ability of AI systems (e.g.,~\cite{Divanji2024}).

\textbf{Challenging} interactions often explore the boundaries of what the AI can and cannot do and correspond to PLEX's \textit{challenge} and \textit{competition} dimensions, supporting how users turned the interaction into a game of skill or knowledge. The \textit{Conquerors} type in BrainHex fits well with this type of interaction, focusing on the enjoyment in competitive play and the satisfaction from beating a well-played opponent. Challenging interactions also correspond to \textit{hard fun} in Lazzaro's Four Fun Keys, emphasizing overcoming obstacles and achieving mastery. Additionally, challenging interactions relate to \textit{ag\^on} in Caillois's patterns of play, involving competitive play. Overall, in the posts and comments, users often expressed a sense of triumph when the AI is ``defeated.'' Thus, these kinds of interactions seemed to unveil the fact that many users, consciously or unconsciously, consider \textbf{AI systems as competitors, rivals, or even enemies}. We worry about their harm and feel accomplished if we can defeat them~\cite{Fischer2023, Acemoglu2021}. The nature of this relationship with AI has been reflected in many domains, such as creative work~\cite{Kawakami2024}, design~\cite{Li2024}, use of smart home technologies~\cite{Felber2022}, and games~\cite{kim16}, to name a few.

\textbf{Tricking} interactions align with PLEX's \textit{subversion} and \textit{control} dimensions, supporting how users attempt to find loopholes or vulnerabilities in the AI's design, either out of curiosity or the thrill of breaking rules. The \textit{Conqueror} and \textit{Achiever} player types in BrainHex fit well with tricking, focusing on overcoming challenges through unconventional means and the desire to complete hidden quests of fully exploiting the AI. Tricking interactions correspond to \textit{hard fun} in Lazzaro's Four Fun Keys, emphasizing the challenge and satisfaction of finding ways to take advantage of the AI's weaknesses. Moreover, tricking can relate to \textit{ag\^on} in Caillois's patterns of play, where users engage in competitive interaction with the AI, trying to outwit it. In general, these interactions revealed that users sometimes treat \textbf{AI systems as something to exploit}, like a type of resource to derive full benefits from. This type of relationship has been dominating studies that use AI technologies to solve practical problems. Some recent works have also explored how users trick AI systems to protect themselves or achieve their goals, such as avoiding personality profiling~\cite{Volkel2020}.

\textbf{Contriving} interactions reflect PLEX's \textit{exploration} and \textit{expression} dimensions, supporting how users prompt the AI to generate novel ideas, stories, or inventions, fostering discovery and innovation. The \textit{Seeker} player type in BrainHex is consistent with contriving, highlighting how users joyed exploring ways to work with AI to develop creative content. Contriving interactions also correspond to \textit{easy fun} in Lazzaro's Four Fun Keys, illustrating the enjoyment users find through carefree creativity and exploration. Additionally, contriving relates to \textit{ilinx} in Caillois's patterns of play, resulting in astonishing surprises from spontaneous and unstructured play. In these ways, the users seemed to consider the \textbf{AI system as a medium or even partner for creative exploration}, working with it to enable unexpected revelations and inspirations. This type of relationship is associated with previous work exploring the role of AI in creativity support~\cite{Candy2020}.

\subsection{Practical Implications}
In addition to allowing us to understand how users assess the AI agency and establish different types of human-AI relationships through play, our framework also has implications for the design of AI-infused systems that we discuss below.

\textbf{Reflecting Interactions:} The details of this category illustrate the potential of AI tools to provide a platform and space for critical thinking. These interactions can be particularly useful in educational contexts, enhancing tools that encourage critical thinking and reflective learning. For instance, AI systems designed for education can engage students in thoughtful discussions, helping them explore different perspectives on a topic. This can be especially beneficial in learning about sensitive and complex subjects like ethics and philosophy, where exploring multiple viewpoints is crucial. Moreover, our results indicated that users tend to inquire about the AI's ``perspective'' on contentious topics about the AI itself, such as defining itself or discussing bias and ethical concerns in AI. These interactions can be leveraged in a way to help users calibrate trust in AI. For example, we found that users asked ChatGPT to reflect on its capacity in decision-making processes, which could be used to enhance explainability in practical decision-making tasks to support establishing trust, as explored in previous work~\cite{Toreini2020, Kim2023}. 

\textbf{Jesting Interactions:} The majority of jesting interactions indicated the users' desire to explore fun and lighthearted materials with AI. This desire and this type of interaction can be leveraged to make AI systems more enjoyable and engaging, both in applications designed for entertainment and serious purposes. For example, incorporating humor and playful interactions can make customer service bots more approachable~\cite{Liu2022}, thereby improving user satisfaction. Previous works have also indicated that humor is valued by users of mental health chatbots~\cite{Wong2024, Wester2024}. The lighthearted interactions we observed could be valuable inspirations for designing chatbot apps to support individuals needing personal, private, and confidential assistants and ``open up'' to the AI assistant, which is important for certain contexts~\cite{Tepper2020Builtin}. Overall, the playful nature of jesting interactions with AI can be used to help make AI systems more inviting and emotionally supportive.

\textbf{Imitating Interactions:} Users' exploration of AI's ability to imitate different personalities and styles has indications of how AI can be used in creative writing contexts~\cite{creativeWr, Gero2023}. For example, AI can adopt various personas and assist writers in generating diverse character dialogues, enhancing the creative process by providing new and varied perspectives. The entertainment industry can also benefit from AI's imitative capabilities. For example, in video game development, AI can create non-playable characters (NPCs) with distinct personalities and dialogue patterns, making the gaming experience more engaging. Moreover, AI's ability to emulate different styles also has educational implications. For example, students can learn by analyzing and comparing AI-generated text that imitates a certain style with masterpieces and/or writings of their own. Language learning tools can also leverage imitating interactions to help learners practice conversational skills by mimicking different types of personality. Furthermore, imitative AI can also play a role in mental health support by adopting a comforting communication style familiar to the user, providing a sense of understanding and companionship. 

\textbf{Challenging Interactions:} This type of interactions revealed a competitive side of human-AI interaction. This may have important implications for game design and gamification. For example, video game design can leverage AI to bring out the competitive side of players, encouraging them to engage in difficult tasks. In educational contexts, this type of interaction can also motivate users to challenge the AI and learn through competition. Moreover, challenging interactions also serve as a way of revealing the limitations of AI and can support education about AI literacy~\cite{Long2020}. For example, learning about how AI output can be incorrect in various scenarios by challenging the AI system can help users establish an appropriate reliance on AI, especially in complex and uncertain tasks in which appropriate reliance is often poorly calibrated~\cite{Salimzadeh2024}.

\textbf{Tricking Interactions:} The level of creativity exhibited by users in outsmarting the AI is particularly intriguing in the tricking interactions. This type of relationship between users and AI can be leveraged to improve the security of sensitive systems. By analyzing how users attempt to deceive AI, developers can identify potential vulnerabilities within the system and use this knowledge to strengthen the system. This approach is particularly important in applications involving sensitive information, such as chatbots for financial services. Additionally, this knowledge can contribute to the development of ethical guidelines and best practices for AI deployment, for example, by adopting methods explored by~\citet{bai2022}. Understanding the creative ways users interact with and attempt to deceive AI also highlights the importance of transparency and robustness in AI systems. Developers can use these insights to create AI that not only resists manipulation but also communicates its limitations and decision-making processes more clearly to users.

\textbf{Contriving Interactions:} It is intriguing to see how users asked ChatGPT to generate things that mankind has not seen yet, such as a new type of color. Interactions like these have important implications in creativity support tools across various fields. For instance, artists can experiment with AI-generated visual concepts, pushing the boundaries of traditional art forms, involving e.g., what Tricaud and Beaudouin-Lafon call an ``epistemic process''~\cite{Tricaud2024}. Similarly, in product design, AI can generate unconventional ideas for new products or improve existing ones by suggesting novel features and functionalities~\cite{Mozaffari2022}. This type of collaborative approach can result in a synergy where human creativity and AI's generative capabilities complement each other, producing results that neither could achieve alone, in what Dellermann et al. call ``hybrid intelligence systems''~\cite{dellermann}. 

\subsection{Limitations and Future Work}
Our study has several limitations that can be addressed in future work. First, we used ChatGPT as a proxy to understand users' playful interactions with the recent advancements of AI. However, this platform only represents one type of AI system -- chat-based natural language interaction. Future studies with other types of AI systems, such as image generators, recommender systems, and personal assistance, need to be conducted to enrich our framework of playful interactions. Second, we used posts on Reddit to understand how users interacted with ChatGPT. While this approach allows us to orient data collection toward playful interactions since Reddit is innately a light-spirited platform, we do acknowledge that there might be playful interactions to be observed in other ways. Future work that leverages direct user studies, such as observations or experiments, may help expose playful interactions unseen by us. Third, because ChatGPT is a newly released system, our study results only captured explorations made by early adopters. We focused on the early adopters intentionally since their experiences provided a unique perspective, exposing interaction patterns when the capability and impact of the AI system are largely unknown. However, further study is needed to explore how users in other phases of the innovation adoption life cycle maintain their relationship with AI (e.g., maintain trust) through play. Finally, our analysis focused on user interactions with ChatGPT before its major update on January 9th, 2023. While this time frame offers valuable insights (as discussed in Section~\ref{sec:methods_collection}), the study might reflect a homogenous set of early interactions, missing the evolution of user behavior over time. Future research should include extended periods to enhance the generalizability of the results.

%% file: s_conclusion.tex
\section{Conclusion}
In this paper, we reported on an empirical analysis to explore users' playful interactions with ChatGPT, an emerging AI system. Our objective was to establish a preliminary framework for characterizing users' playful interactions with AI systems, utilizing its lexicon as a lens to uncover key factors contributing to human-AI relationships. Through analyzing 372 posts on the ChatGPT subreddit, we found that more than half of the posts were exhibiting playful interactions. A thematic analysis revealed six main categories encompassing various subcategories of these interactions. The main categories were: reflecting, jesting, imitating, challenging, tricking, and contriving. By triangulating our findings with established frameworks of agency and gameplay experiences, we offered insights into how users used playful interactions to assess the AI agency and how this type of interaction revealed various facets of the human-AI relationship. We reflected on how previous works on human-AI interaction are related to each category of playful interaction identified in our framework and their practical implications for the design of AI-infused systems. Overall, our work contributes to a niche body of literature by providing empirical insights into how playful interactions provide a window into and influence the way in which humans perceive and navigate the ever-evolving landscape of new technologies.